\documentclass[prl,amsmath,superscriptaddress,twocolumn,longbibliography]{revtex4-1}

\usepackage{epsfig}
\usepackage{hyperref}
\usepackage{amssymb}
\usepackage{amsmath}
\usepackage{amsfonts}
\usepackage{soul}    % use /st{} to strikeout some text
\usepackage{bm}
\usepackage{braket}
\usepackage{graphicx}
\usepackage{xcolor}
\usepackage{float}
\usepackage{lipsum}
\usepackage{comment}
\usepackage{siunitx}
\newcommand{\Jnatphys}{Nat. Phys.}

\newcommand{\Jprl}{Phys. Rev. Lett.}

\newcommand{\Jpra}{Phys. Rev. A}
\newcommand{\Jprb}{Phys. Rev. B}

\newcommand{\Jepl}{Europhys. Lett.}

\newcommand{\Jepjd}{Eur. Phys. J. D}

\newcommand{\crasphy}{C. R. Phys.}

\newcommand{\JRepProgPhys}{Rep. Prog. Phys.}

\newcommand{\JAnnualRevCondMat}{Annual Rev. Cond. Mat. Phys.}

\renewcommand{\d}{\ensuremath{\mathrm{d}}}
\newcommand{\e}{\ensuremath{\mathrm{e}}}
\newcommand\avg[1]{\langle #1 \rangle}

\begin{document}
\title{Quench Spectroscopy of a Disordered Quantum System}

\author{L. Villa}
\email{louis.villa@polytechnique.edu (he/him/his)}
\affiliation{CPHT, CNRS, Ecole Polytechnique, IP Paris, F-91128 Palaiseau, France}
\author{S. J. Thomson}
\email{steven.thomson@polytechnique.edu (he/him/his)}
\affiliation{CPHT, CNRS, Ecole Polytechnique, IP Paris, F-91128 Palaiseau, France}
\affiliation{JEIP, USR 3573 CNRS, Coll\`ege de France, PSL Research University, 11 Place Marcelin Berthelot, 75321 Paris Cedex 05, France}
\author{L. Sanchez-Palencia}
\email{laurent.sanchez-palencia@polytechnique.edu (he/him/his)}
\affiliation{CPHT, CNRS, Ecole Polytechnique, IP Paris, F-91128 Palaiseau, France}
\date{\today} 

\begin{abstract}
The characterization of excitations in disordered quantum systems is a central issue in connection with glass physics and many-body localization.
Here, we show that quench spectroscopy of a disordered model, as realized from its out-of-equilibrium dynamics following a global quench,
allows us to fully characterize the spectral properties of the disordered phases.
In the Bose-Hubbard model, a clear signature of gapless excitations in momentum-resolved spectroscopy enables us to accurately locate the Mott insulator to Bose glass transition,
while the presence or absence of a well-defined soundlike mode distinguishes the superfluid from the Bose glass phase.
Moreover, spatially-resolved spectroscopy provides local spectral properties and allows us to extract the typical spacing of gapless regions, giving a second independent way to uniquely identify all three phases.
Our findings have far-ranging implications for a variety of experimental platforms,
and offer a powerful and versatile probe of the low-energy phases of disordered systems.
\end{abstract}

\maketitle

Understanding the interplay of disorder and many-body interactions in quantum matter is a longstanding problem which remains a highly active area of modern research. While low-dimensional non-interacting systems can be completely localized by even an infinitesimal concentration of disorder~\cite{Anderson58,abrahams1979}, interacting systems exhibit much richer behavior~\cite{Fleishman+80}, from exotic quantum glass ground states~\cite{Giamarchi+88,Fisher+89,krauth1991,rapsch1999}
to collective Anderson~\cite{gurarie2002,gurarie2003,bilas2006,lugan2007b,lugan2011,lellouch2015} and
many-body~\cite{Basko+06,oganesyan2007,Huse+13,altman2015,alet2018,AbaninEtAlRMP19} localization. 
One particularly pressing question is how excitations from glassy ground states behave in light of recent suggestions that quantum glass ground states may be smoothly connected to many-body localization of highly excited states~\cite{AbaninEtAlRMP19}. With this question in mind, it is highly desirable to develop methods to probe the excitations of quantum glasses.
This is traditionally realized using Bragg spectroscopy~\cite{fallani2007,roux2013Dynamic} but it remains an extremely challenging task,
which requires fine tuning of both the momentum and the frequency probed.
Moreover, it is not suitable for probing local properties of inhomogeneous systems.

Recent advances on the control of strongly interacting quantum matter,
with possibly single-site resolution imaging in optical lattices~\cite{Bakr+09,Sherson+10,Haller+15,Cheuk+15,Parsons+15,Omran+15,Edge+15}, allow us to reconsider these issues from the perspective of out-of-equilibrium dynamics.
A first step in this direction has been reported in Ref.~\cite{meldgin2016} where the Bose glass to superfluid transition has been identified via the proliferation of fluctuations following a quench across the transition. 
Recently, quenches have been used to probe localization of highly excited states and identify many-body mobility edges~\cite{RYao+20}.
Here we develop a new form of excited-state quench spectroscopy which can provide either momentum or spatially resolved information of disordered quantum systems
and permits us to fully characterize the spectral properties of the zero temperature quantum phases . This should be distinguished from prior work on quench spectroscopy of homogeneous models~\cite{villa2019Unraveling,villa2020Local}, as the explicit breaking of translation invariance in the Hamiltonian here leads to markedly different behavior.

In order to provide a proof-of-concept demonstration of our approach, we benchmark it using one-dimensional (1D) disordered bosons,
where exact numerical calculations can be performed.
We show that quench spectroscopy provides all necessary information to characterize the excitation spectra and determine the phase diagram.
Moreover, we obtain valuable local spectral properties, including the real-space distribution of gapped and gapless regions in the Bose glass phase.
Our approach may be implemented in present-day quantum simulators and permits full characterization of the quantum phases of disordered bosons, including the still elusive Bose glass.
Extensions of quench spectroscopy to higher dimensions and other disordered systems is discussed.

\emph{Model} -
\label{sec.model}
Interacting bosons in a disordered potential may be described by the disordered Bose-Hubbard model (DBHM), the 1D Hamiltonian of which reads as
\begin{equation}
\label{eq:hamiltonian_bhmd}
\hat{H}=\sum_{j}\left[-J\left(\hat{a}_{j}^{\dagger}\hat{a}_{j+1}+\text{h.c.}\right)+\frac{U}{2} \hat{n}_j (\hat{n}_j - 1) + V_j\hat{n}_{j}\right],
\end{equation}
where $\hat{a}_j$ and $\hat{a}^{\dagger}_{j}$ are, respectively, the annihilation and creation operators of a boson on site $j$,
$V_j = \Delta_{j}-\mu$,
with $\Delta_j$ a site-dependent random potential
and $\mu$ the chemical potential.
The applications of the DBHM range from disordered superfluid Helium~\cite{Fisher+89} to magnetic systems~\cite{Matsubara+56,Giamarchi+08,Yu+10,Zheludev+13,Zapf+14,Thomson+15}.
It has also been emulated in ultracold-atom systems~\cite{lsp2010,modugno2010},
where the disorder can be generated by a speckle pattern~\cite{clement2006Experimental,White+09},
a bichromatic quasiperiodic potential~\cite{derrico2014,gori2016},
impurities~\cite{gavish2005,paredes2005,gadway2011}
or by a spatial light modulator~\cite{Choi+16,Bruce+15}.
Hereafter, we consider a disordered potential drawn from a box distribution $\Delta_j \in \left[-{\Delta}/{2},{\Delta}/{2}\right]$
with $\Delta$ the disorder strength.

\begin{figure}[t!]
\begin{center}
\includegraphics[width=\columnwidth]
{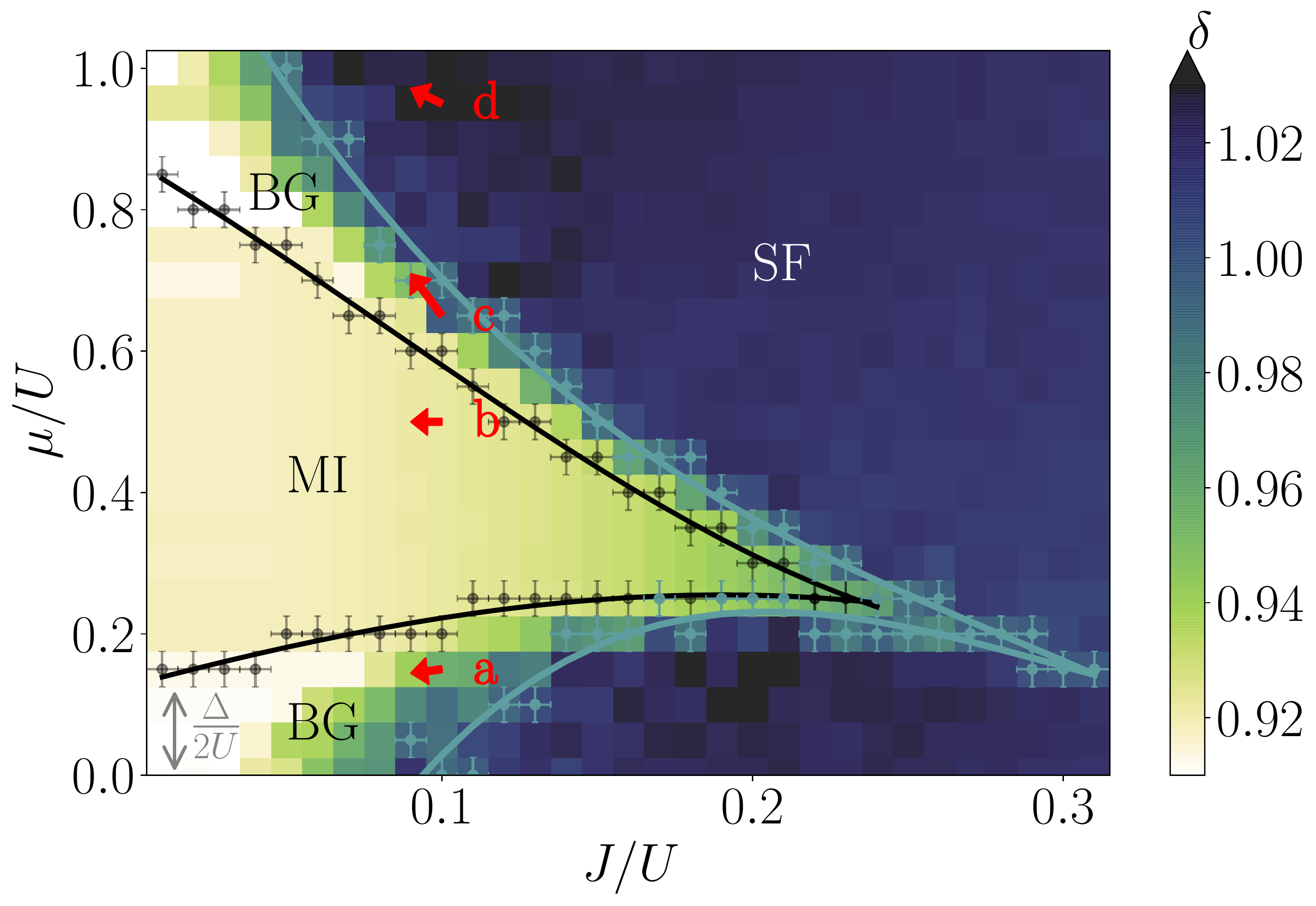}
\caption{Phase diagram of the 1D DBHM computed using DMRG ($L=47$, $N_{\rm{s}}=15$) at $\Delta/U=0.25$.
The transition points are found by analyzing the compressibility $\kappa$ and the one-body correlator, and the error bars reflect the size of the underlying grid used to compute the boundaries.
The gray points show the transition from $\kappa=0$ (MI phase) to $\kappa \neq 0$, to which the solid black line is fitted.
The parameter $\delta$ is plotted in color scale. The cyan points where $\delta=1$, indicating the transition between the BG and SF phases, to which the cyan solid line is fitted.
The red arrows represent the quenches of $J/U$ along lines of constant density
as considered in Fig.~\ref{fig.qsf_random}.
}
\label{fig.phase}
\end{center}
\end{figure}

The quantum phase diagram of the 1D DBHM has been extensively studied previously~\cite{scalettar1991,prokofev1998Comment,rapsch1999Density,yao2016Superfluid}.
For reference, we reproduce it in Fig.~\ref{fig.phase} using density matrix renormalisation group (DMRG) simulations~\cite{schollwock2011Densitymatrix} for the system size $L=47$ as used throughout this work, averaged over $N_{\rm{s}}=15$ disorder realizations.
The equilibrium, zero temperature phase diagram for the clean system ($\Delta=0$) contains two phases: a gapped incompressible Mott insulator (MI) and a gapless compressible superfluid (SF). 
When disorder is added into the model, a third gapless compressible phase intervenes between the other two~\cite{Pollet+09,Gurarie+09},
known as the Bose glass (BG)~\cite{Giamarchi+88,Fisher+89}.
To identify those three phases, we compute the compressibility $\kappa=\partial n/\partial \mu$ and the one-body correlator $g_1(i,j) = \langle \hat{a}^{\dagger}_i \hat{a}_j \rangle$:
The compressibility allows us to distinguish the MI (which is the only incompressible phase, $\kappa=0$), from the other two ($\kappa \neq 0$).
The one-body correlator
decays exponentially with the distance $r=|i-j|$ in both the MI and BG phases, and algebraically in the SF. 
We introduce the relative Pearson's coefficient $\delta$, which provides a sensitive probe of the relative quality of exponential and algebraic fits.
By identifying the point at which $g_1(r)$ crosses over from exponential ($\delta < 1$) to algebraic ($\delta > 1$) decay, we obtain a good estimate of the BG-SF transition, see Refs.~\cite{Yao+20,SM}.

\begin{center}
\begin{figure}[t!]
\includegraphics[width=\linewidth]{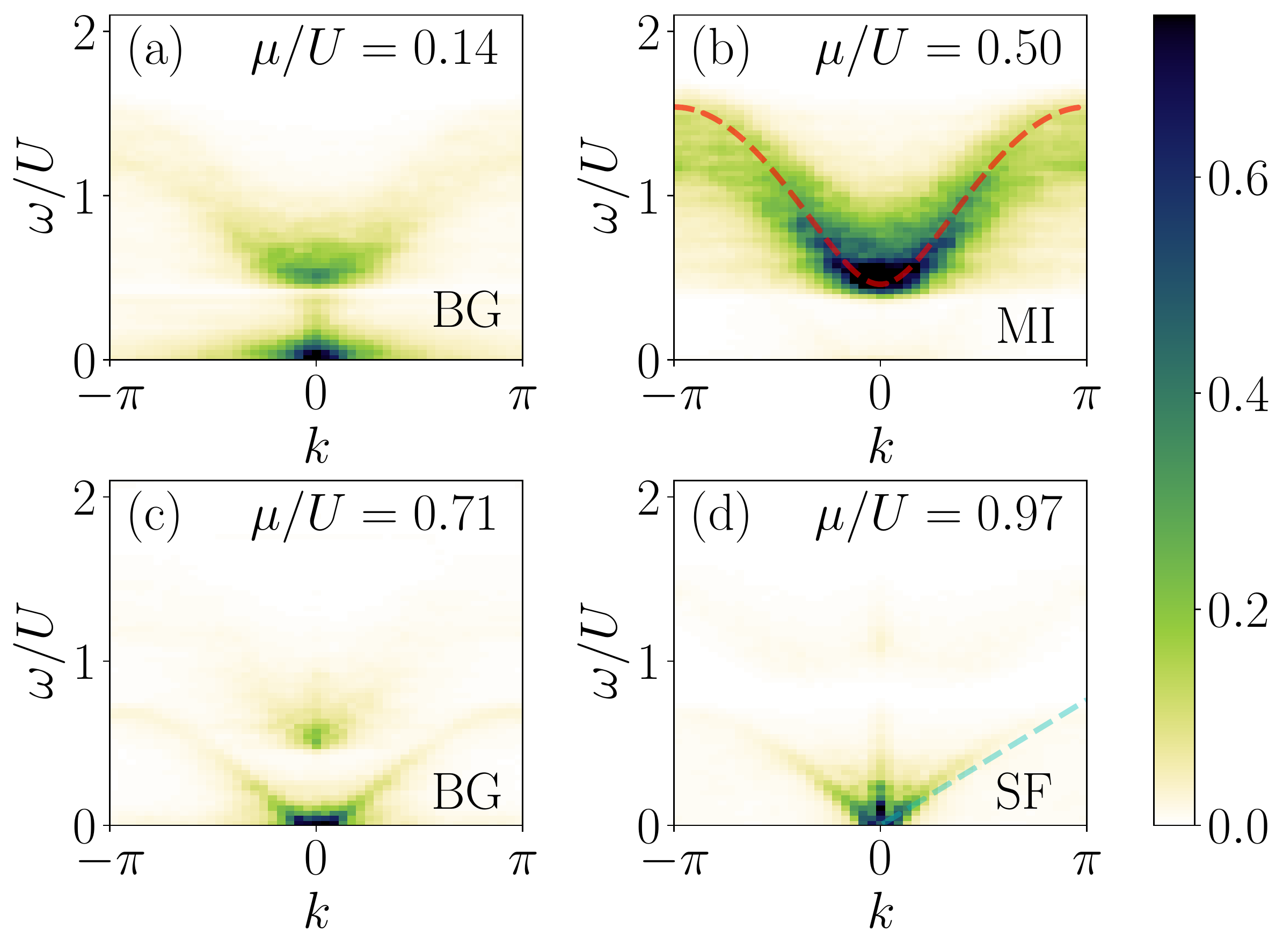}
\caption{QSF of the $g_1$ correlator ($L=47$, $N_{\rm{s}}=25$) at $\Delta/U=0.25$, after a quench from $J_{\rm{i}}/U = 0.1$ to $J_{\rm{f}}/U=0.09$, and various values of $\mu/U$ (see red arrows in Fig.~\ref{fig.phase}).
(a)~and (c)~correspond to the BG phase at $\mu/U=0.14$ and $\mu/U=0.71$, respectively,
(b)~to the MI phase at $\mu/U=0.5$, and (d)~to the SF phase at $\mu/U=0.98$.
The~dashed red line in panel~(b) is the excitation band of the non-disordered MI.
The dashed cyan line in panel~(d) is a linear fit to the QSF close to $k=0$ in the SF phase.
}
\label{fig.qsf_random}
\end{figure}
\end{center}

\vspace{-10mm}
\emph{Out-of-equilibrium dynamics} -
\label{sec.qsf}
To induce out-of-equilibrium dynamics, we
first prepare the system in its ground state using DMRG with given values of $\mu/U,J/U$, and $\Delta/U$.
We then quench the hopping from $J_{\rm{i}}/U=0.1$ to $J_{\rm{f}}/U= 0.09$ at a fixed density, hence also changing $\mu$, see red arrows in Fig.~\ref{fig.phase}.
The state then evolves out-of-equilibrium under the unitary dynamics generated by the new Hamiltonian, computed using the many-body time-dependent variational principle (TDVP)~\cite{haegeman2016Unifying}, using the hybrid time evolution method~\cite{Goto+19,Paeckel+19,Chandra+20}. In all of the following, we use open boundary conditions, a maximum bond dimension $\chi=128$ and a maximum evolution time of $t_{\rm{max}}=20 /J_{\rm{i}}$, with timesteps $\d t=0.01/J_{\rm{i}}$, and we truncate the local occupancy to a maximum of $N_{\rm{b}}=5$ bosons per site. We have checked these parameters and found them to yield well-converged results. For further details, see Ref.~\cite{Paper2}. The timescale and system size are consistent with state-of-the-art current experiments~\cite{trotzky2012Probing}.

The out-of-equilibrium dynamics of an observable $\hat{O}(x,t)$, at any time $t$ after the global quench and a distance $x$ in real space, is given by
\begin{equation}
\label{eq:observable}
G(x,t) = \avg{\hat{O}(x,t)}=\text{Tr}\big[\hat{\rho}_{\textrm{i}}\,\hat{O}(x,t)\big],
\end{equation}
where $\hat{\rho}_{\rm{i}}$ is the density matrix of the initial state.
Spectral properties of the excitations may then be obtained using the space-time Fourier transform of Eq.~\eqref{eq:observable} (quench spectral function, QSF),
\begin{equation}
\label{eq:QSF_disorder_full_expression}
\begin{split}
G(k,\omega)&=2\pi\int\d x\,\e^{-ikx}\sum_{n,n'}\delta(E_{n'}-E_{n}-\omega)\\
&\quad\times\rho_{\rm{i}}^{n'n} \bra{n}\hat{O}(x)\ket{n'},
\end{split}
\end{equation}
where $\vert n\rangle$ represents the many-body states of the model, of energy $E_n$.
Weak quenches populate the low-lying excited states and,
as previously shown, different properties of the excitations can be measured from the QSF, depending on the choice of the observable~\cite{villa2019Unraveling,villa2020Local}. In contrast with these prior works, however,
the disorder breaks translation invariance and the energy eigenstates do not have a well-defined momentum $k$. Moreover,
a global quench of a disordered system generates
single-particle excitations, which are forbidden by translation invariance in homogeneous systems.
Hence the application of quench spectroscopy to disordered systems is not trivial and requires some care to interpret.
In clean systems, the dispersion relation $E_k$ of the excitations have been obtained in both the MI and the SF using the one-body correlator $\langle\hat{O}(x,t)\rangle=g_1(x,t)=\langle\hat{a}^{\dagger}(x,t) \hat{a}(0,t)\rangle$.
The latter may be experimentally measured via standard time-of-flight imaging~\cite{Greiner+01}.
In the DBHM studied here, although the energy eigenstates do not have a well-defined momentum $k$,
the Fourier transform in Eq.~(\ref{eq:QSF_disorder_full_expression}) remains well-defined and
weak disorder only broadens the spectral features~\cite{Paper2}.
Below, we show that the observable $\langle\hat{O}(x,t)\rangle=g_1(x,t)$ provides distinguishing characteristics of all three phases of the DBHM using the quench spectroscopy protocol described above. 

%\vspace{-8mm}
\emph{Numerical results} - 
\label{sec.results}
Figure~\ref{fig.qsf_random} shows the modulus of the QSF, $\vert G(k,\omega) \vert$, of the observable $g_{1}(x,t)$
for four different choices of $\mu/U$ spanning the three phases.
The quenches performed are indicated by red arrows in Fig.~\ref{fig.phase}.
For all data, we use a Hann window function to reduce boundary effects before taking a Fourier transform, and we subtract the long-time average.
The results are then averaged over $N_{\rm{s}}=25$ disorder realizations
and normalized.
In all cases, the low-$k$ behavior contributes more strongly to the observed signal.
This is because of significant scattering of excitations from the disorder on small length scales (large $k$), which broadens the spectrum at large momenta and results in a weak signal.
In contrast, scattering is expected to be screened by repulsive interactions at low momenta, hence resulting in a stronger signal, weakly affected by the disorder~\cite{lugan2007b,lugan2011}.

\begin{center}
\begin{figure}[t]
\includegraphics[width=\linewidth]{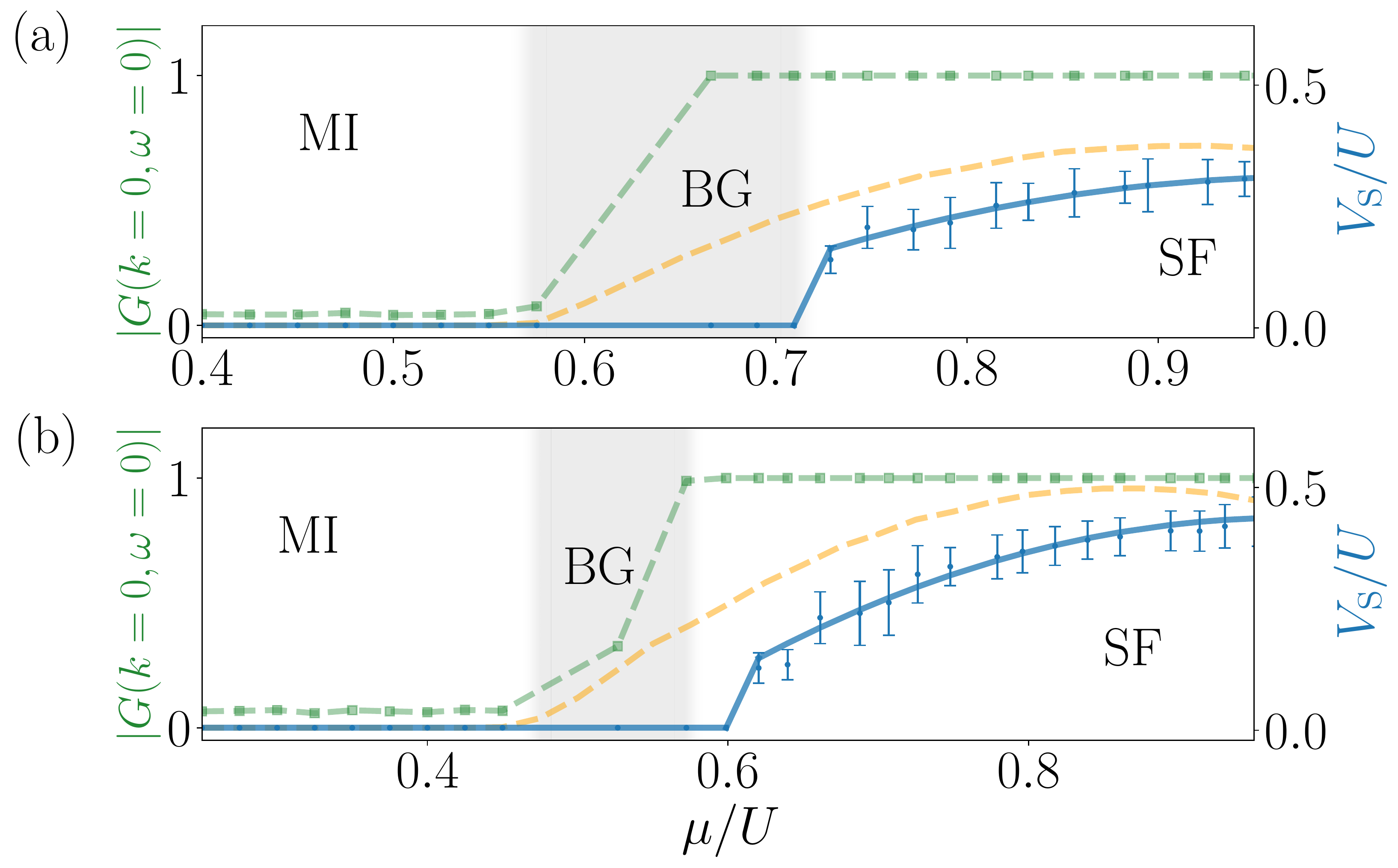}
\caption{
Identification of the three phases of the DBHM from the amplitude of the normalized QSF of $g_1(x,t)$ at the origin ($\vert G(k=0,\omega=0)\vert$, green squares) and the speed of sound ($V_{\rm{s}}$, blue circles), for (a)~$J_{\rm{f}}/U=0.09$, and (b)~$J_{\rm{f}}/U=0.12$, averaged over $N_{\rm{s}}=25$ disorder realizations.
The solid blue line is a piecewise quadratic fit intended as a guide to the eye.
The MI is characterized by $\vert G(0,0)\vert=0$ and $V_{\rm{s}}=0$,
the BG by $\vert G(0,0)\vert \neq 0$ and $V_{\rm{s}}=0$,
and the SF by $\vert G(0,0)\vert \neq 0$ and $V_{\rm{s}} \neq 0$.
The dashed orange line shows the speed of sound obtained from the mapping of the clean model to spinless fermions.
The grey region represents the BG phase as identified by the data in Fig.~\ref{fig.phase}.
}
\label{fig.speed_of_sound}
\end{figure}
\end{center}

\emph{Characterizing disordered phases using QSF} - 
We now discuss how to characterize the three phases expected in the DBHM from the QSF data.
The BG and SF can both be distinguished from the MI by the existence or absence of gapless excitations. 
As expected, in the MI, the resonances of the QSF are strongest around $\omega/U \sim 1$ and, most importantly for our purposes, there is no signal close to $\omega/U=0$, hence signaling a finite gap.
More precisely, the spectrum measured by the QSF
closely matches the corresponding excitation spectrum for the clean system in the MI with $\bar{n}=1$, $E(k)=\sqrt{(U-6J_{\rm{f}}\cos k)^2+32(J_{\rm{f}}\sin k)^2}$~\cite{barmettler2012Propagation}, see dashed red line in Fig.~\ref{fig.qsf_random}(b).
In spite of significant disorder-induced broadening of the QSF, the gap, $\varepsilon \simeq U-6J_{\rm{f}}$,
is almost unaffected by the disorder, owing to strong screening in the low $k$ sector. This value is to be contrasted to the expected gap $\varepsilon=U-\Delta$ in the atomic limit ($J_{\rm{f}}=0$).
By contrast, in both the BG and SF phases, we find a strong peak in the QSF close to zero frequency, indicating the existence of gapless excitations. 
By extracting the QSF amplitude at $\omega=0$ and $k=0$,
we thus clearly distinguish the gapped MI ($\vert G(k=0,\omega=0)\vert = 0$) from the gapless BG and SF ($\vert G(k=0,\omega=0)\vert \neq 0$) phases.
Numerical results for two different values of $J_{\rm{f}}/U$ are shown in Fig.~\ref{fig.speed_of_sound} (green points and dashed line).
The onset of gapless excitations measured by $\vert G(0,0)\vert$ matches well with the MI-BG phase transition in Fig.~\ref{fig.phase} (left boundary of the gray region in Fig.~\ref{fig.speed_of_sound}).

To distinguish the BG and SF phases, we use the qualitatively different behaviors exhibited by the QSF due to the different natures of their low-lying excitations.
In the SF [Fig.~\ref{fig.qsf_random}(d)], the QSF shows a clear V-shaped continuum emerging from the origin,
whereas in the BG [Fig.~\ref{fig.qsf_random}(a) and (c)] the QSF is featureless close to $\omega/U=0$.
This may be attributed to the existence of a well-defined speed of sound in the SF, which is absent in the BG.
We can then discriminate the SF and BG phases by the presence or absence of a soundlike mode with finite velocity.
The velocity of the latter is numerically extracted by a linear fit close to the origin of the (disorder-averaged) QSF, the results of which are shown in Fig.~\ref{fig.speed_of_sound} (blue points and solid line).
To extract the slope, we perform several linear fits across different momentum intervals $k \in [0,k_{\textrm{max}}]$ while varying $k_{\textrm{max}}$.
The error bars in Fig.~\ref{fig.speed_of_sound} are given by the standard deviation of the sound velocities obtained by these different fits~\cite{SM}.
As shown in Fig.~\ref{fig.speed_of_sound}, the results exhibit a clear SF-BG transition, which closely matches that in Fig.~\ref{fig.phase}
(right boundary of the gray region in Fig.~\ref{fig.speed_of_sound}).

\vspace{-10mm}
To estimate the speed of sound, we may map the DBHM onto spinless fermions in the interaction-dominated regime~\cite{cazalilla2003Onedimensional,cazalilla2004Differences}. Since the soundlike mode is relevant in the low $k$ limit, we may further neglect the disorder. In the regime with average boson filling $1<\overline{n}<2$ corresponding to Fig.~\ref{fig.speed_of_sound},
strongly-interacting bosons moving on top of a uniformly filled `vacuum' map onto free fermions with the average density $\overline{n}_{\rm{f}} = \overline{n}-1$.
The speed of sound then maps onto the Fermi velocity with a factor of $2$ due to Bose enhancement.
It yields $V_{\rm{s}} = 4J \sin(\pi \overline{n}_{\rm{f}})$, shown as dashed orange lines in Fig.~\ref{fig.speed_of_sound}.
This result is in remarkable agreement with the fitted velocities within the SF phase for the non-disordered model (data not shown here, see Ref.~\cite{SM}).
In the presence of disorder, we find that the speed of sound is renormalized towards lower values, as expected from renormalization group analysis within Luttinger liquid theory~\cite{Giamarchi+88}. Here we find that this renormalization is weak in the SF phase, down to the SF-BG transition where $V_{\rm{s}}$ drops to zero.

Hence the QSF displays clear features that allow us to \emph{quantitatively} distinguish the three phases:
The amplitude of the QSF at $\omega=0, k=0$ discriminates the gapped phase (MI) from the gapless phases (SF and BG),
and a well-defined soundlike mode uniquely identifies the SF.
The complete phase diagram of the DBHM can be systematically reconstructed by analysis of the QSF~\cite{Paper2}.

\begin{center}
\begin{figure}[t]
\includegraphics[width= \linewidth]{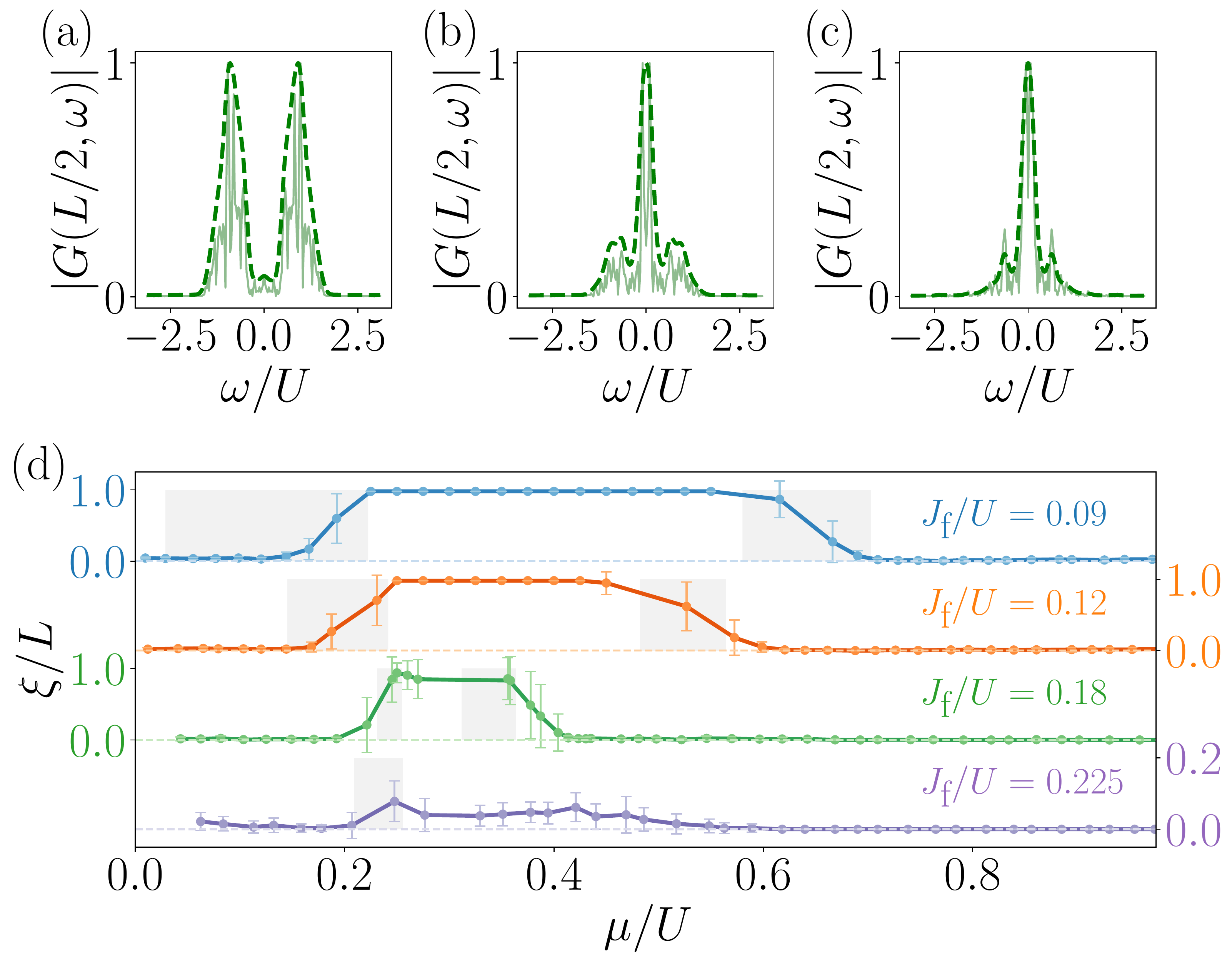}
\caption{
Spatially-resolved quench spectroscopy.
(a)-(c)~Normalized LSF $|G(x=L/2,\omega)|$ on the center lattice site for a single disorder realization, a quench from $J_{\rm{i}}/U=0.1$ to $J_{\rm{f}}/U=0.09$,
and three values of $\mu/U = 0.525, 0.695, 0.820$ in the MI, BG, and SF phases respectively. The dashed line is the result of Gaussian smoothing. 
(d)~SF region spacing $\xi$ versus $\mu/U$ averaged over $N_{\rm{s}}=25$ samples for four values of $J_{\rm{f}}/U$, offset for clarity.
The standard deviation of $\xi/L$ across disorder realizations is indicated by the error bars.
The grey regions represent the BG phase as identified in Fig.~\ref{fig.phase}.
}
\label{fig.LSF}
\end{figure}
\end{center}

\vspace{-9mm}
\emph{Spatially-resolved quench spectroscopy} - 
In the BG, the QSF shows clear signals of independent bands of gapped and gapless excitations [Fig.~\ref{fig.qsf_random}(a) and (c)], a potential indicator of the coexistence of MI and SF regions within the same sample. To get further insight into the real-space distribution of gapped/gapless regions within each realization, we introduce the \emph{local spectral function} (LSF)
\begin{equation}
G(x,\omega)=2\pi\sum_{n,n'}\rho_{\rm{i}}^{n'n} \delta(E_{n'} - E_{n} - \omega) \braket{n | \hat{O}(x) | n'}.
\end{equation}
By identifying whether $G(x,\omega)$ exhibits a peak at $\omega=0$, we may associate a lattice site $x$ with gapped or gapless excitations.
The simplest possible choice to achieve this is to consider density fluctuations $\hat{O}(x,t)=\delta \hat{n}(x,t) = \hat{n}(x,t)-\mathcal{N}$,
where $\mathcal{N}$
is the long-time-average of $\hat{n}(x,t)$.
This quantity can be experimentally measured using quantum gas microscopes with single-site resolution~\cite{Bakr+09,Sherson+10,Haller+15,Cheuk+15,Parsons+15,Omran+15,Edge+15}.
Intuitively, density fluctuations are related to the propagation of excitations in the system, so that the LSF of $\delta\hat{n}(x,t)$ gives a gapless response in both the BG and SF phases, and is gapped in the MI~\cite{Paper2}. This spatially resolved probe following a global quench in an inhomogeneous system should not be confused with prior work
on local quench spectroscopy~\cite{villa2020Local}, which instead focused on the dynamics of a translation-invariant system following a local quench.

Representative results are shown in Fig.~\ref{fig.LSF}(a)-(c). We use a Gaussian convolution to smooth the signal in frequency before extracting the excitation peak~\cite{SM}.
In the MI [Fig.~\ref{fig.LSF}(a)], we find two clear peaks close to $\omega = \pm U$ while, in the SF [Fig.~\ref{fig.LSF}(c)], we find a single peak at $\omega=0$.
In the BG [Fig.~\ref{fig.LSF}(b)], we find both these two features, suggesting the coexistence of multiple types of excitations.
By computing $G(x,\omega)$ for each lattice site, we extract the typical size $\xi$ of gapped regions within a single disorder realization, with results as shown in Fig.~\ref{fig.LSF}(d).
The MI is characterized by a single gapped region with $\xi=L$ while the SF is characterized by the absence of gapped regions, e.g., $\xi \simeq 0$.
In the BG, $\xi/L$ takes on intermediate values, which continuously grow from SF to MI.
As shown in Fig.~\ref{fig.LSF}(d), the three phases identified using this probe closely match those found in Fig.~\ref{fig.phase}.
This lengthscale hence provides a single observable able to discriminate the three phases and enables full reconstruction of the phase diagram~\cite{Paper2}.
Moreover, it provides valuable information on the \emph{distribution} of SF regions within the BG, a key quantity for determining many properties of this phase. For instance, it allows direct measurement of the growth of SF regions, and could allow for direct observation of the percolation transition (in $d>1$) from the BG to the SF phase~\cite{Niederle+13}.

\emph{Discussion/Conclusion} -
\label{sec.conclusion}
In this work, we have demonstrated that quench spectroscopy accomplishes two key goals in the DBHM.
Firstly, momentum resolved quench spectroscopy is capable of distinguishing all three phases of the model by testing whether the excitations are gapped or gapless, and in the latter case whether or not they exhibit a soundlike mode characteristic of superfluidity.
It provides complementary information on the energy-momentum profile of excitations,
similar to standard Bragg spectroscopy~\cite{stoeferle2004,fallani2007,roux2013Dynamic} but beyond lattice shift and modulation spectroscopy~\cite{Mun+07,Schachenmayer+10,orso2009Lattice,citro2020Lattice} and using a greatly simplified experimental protocol~\cite{villa2019Unraveling,villa2020Local}.
Secondly, we have introduced spatially-resolved quench spectroscopy, particularly fruitful in inhomogenous systems. Local spectral properties allow identification of the distribution of gapped and gapless regions, the typical size of which provides a single parameter to distinguish the three phases.
Both protocols can be used to systematically map out the entire phase diagram~\cite{Paper2}.
It is one of the aims of this work to stimulate experiments realizing quantum simulators for the disordered Hubbard model or other disordered models. It is expected that the quench spectroscopy approaches we propose here will provide experimentalists with an accurate probe relatively easily implemented in such experiments.

The use of quench spectroscopy for disordered systems extends beyond the 1D DBHM model to higher dimensions~\cite{Choi+16},
disordered fermions~\cite{Schreiber+15}, and spin models~\cite{smith2016Manybody}. It also applies to continuous models, recently considered as good candidates to observe the still elusive BG phase~\cite{viebahn2019,yao2019,Yao+20,sbroscia2020,gautier2021}. The extension to higher-dimensional systems is particularly promising, as this is a regime for which efficient numerical methods are scarce, but the experimental realization is comparatively straighforward. Our work also paves the way to detailed experimental investigations into rare-region Griffiths effects in disordered systems, an increasingly important question not only for BG physics but many-body localization~\cite{AbaninEtAlRMP19}, where a different form of quench spectroscopy has already been used to identify mobility edges~\cite{RYao+20} and spectral functions of local operators have been used to study the effects of weak system-bath coupling~\cite{Nandkishore+14}.

\acknowledgments
Numerical calculations were performed using HPC resources from CPHT
and GENCI-CINES (Grants 2019-A0070510300 and 2020-A0090510300).
We acknowledge use of the QuSpin~\cite{weinberg2017quspin,weinberg2019quspin} and TenPy~\cite{Hauschild+18} packages.

\bibliographystyle{apsrev4-1}
\bibliography{refs,refs2,biblioLSP}

\clearpage
\onecolumngrid

\section{Supplemental Material}

\subsection{Computing the phase diagram using standard probes}

The phase diagram shown in Fig. 1 of the main text was computed using the density matrix renormalization group (DMRG) method, with system size $L=47$ and bond dimension $\chi=128$. The MI-BG phase boundary was computed from the points at which the density deviates from an average filling of $\overline{n}=1$ (which coincide with the points where the reduced compressibility $\kappa = \partial n/\partial \mu$ deviates from zero). The BG-SF phase boundary was computed from the decay of the one-body correlator $g_1(i,j)= \langle \hat{a}^{\dagger}_i \hat{a}_j \rangle$, which we discuss here in detail.

\begin{center}
\begin{figure}[h!]
\includegraphics[width=\linewidth]{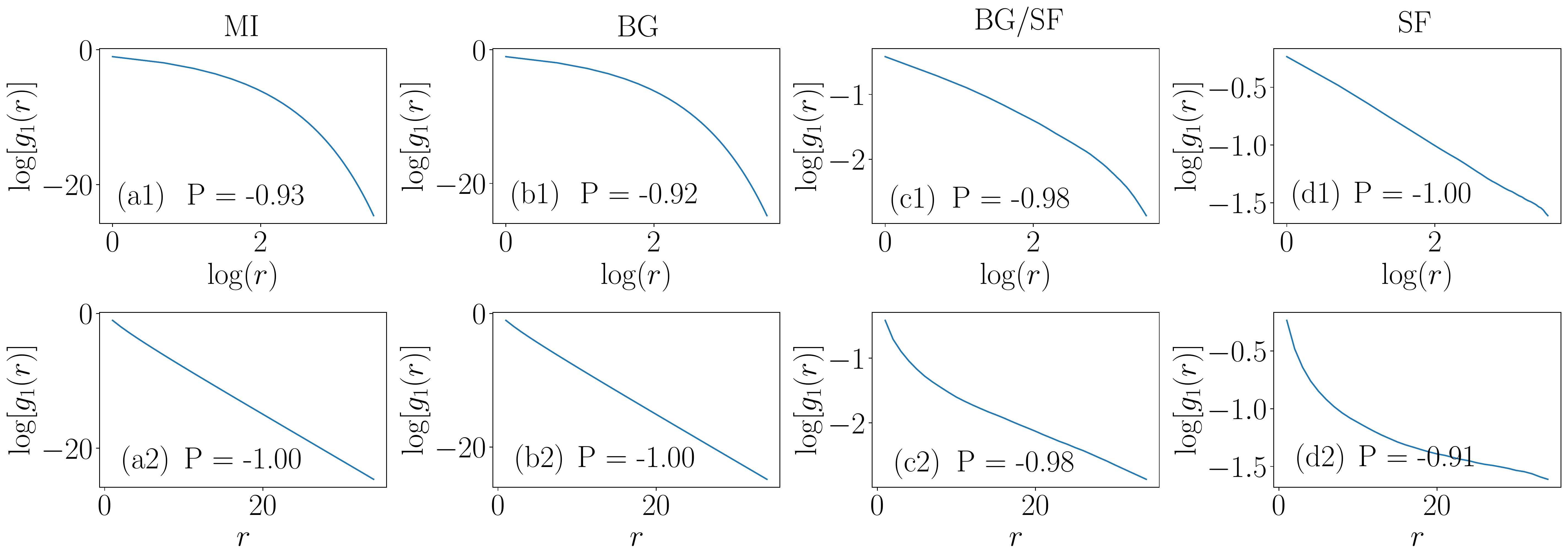}
\caption{Representative plots of $g_1(r)$ in each of the three phases, shown in log-log scale (top row) and semi-log scale (bottom row), for $J/U=0.1$ and chemical potentials
(a)~$\mu/U=0.5$,
(b)~$\mu/U=0.6$,
(c)~$\mu/U=0.7$, and
(d)~$\mu/U=0.8$. In the MI and BG phases, the data are linear on a semi-log fit (with $P^{\textrm{semi-log}} \sim -1$), corresponding to exponential decay. In the SF, the data is closer to linear on a log-log fit (with $P^{\textrm{power-law}} \sim -1$), corresponding to an algebraic decay.
In panel~(c), both fits are of similar quality, indicating close proximity to the transition.}
\label{fig.g1}
\end{figure}
\end{center}

The SF is typically distinguished from the BG via the presence of a non-zero superfluid stiffness in the thermodynamic limit. However, computing this quantity is numerically challenging using tensor network methods. 
An alternative option, which we use here, is to instead study the form of the one-body correlator. In the MI and BG phases, $g_1(i,j)$ decays exponentially with distance $r=|i-j|$. In the SF, $g_1(i,j)$ instead exhibits an algebraic decay $g_1(i,j) \propto |i-j|^{1/2K}$ where $K$ is the Luttinger parameter. In a translationally invariant system, $g_1(i,j) \equiv g_1(|i-j|)$. In contrast, for disordered system one must average over all fixed distances $r=|i-j|$ in order to obtain an average $g_1(r)$, the form of which defines whether the system is either in the MI or BG phase or in the SF phase. Representative plots are shown in Fig.~\ref{fig.g1} in log (top row) and semi-log (bottom row) scales.

In order to determine whether $g_1(r)$ decays algebraically or exponentially, we compute the Pearson correlation coefficient $P$ for a linear fit of $g_1(r)$. This quantity tests for a linear correlation in a given sample: for a perfect linear correlation, $P=1$, and for a perfect linear anticorrelation (as in the case of a decaying function), $P=-1$. By fitting the data in both semi-log and log-log scales, we may obtain a measure of whether $g_1(r)$ is best fitted by an exponential decay ($P \sim -1$ for the data in a semi-log scale) or a power-law ($P \sim -1$ for the data in log-log scale), following Ref.~\cite{Yao+20}. To obtain the phase diagram, we define the ratio of the two coefficients $P$ as
\begin{align}
\delta = \frac{P^{\textrm{power-law}}}{P^{\textrm{semi-log}}},
\label{eq.pearson}
\end{align}
and the phase boundary between BG and SF phases is given by the point at which $\delta=1$. Note that to avoid finite size effects, we restrict the computation of $P$ to the points where $r \ll L$.

\clearpage
\subsection{Extracting the speed of sound}

The group velocity of excitations in a lattice superfluid is given by the slope of the dispersion relation close to the origin, $\textrm{d}E_k/\textrm{d}k$. For phonons with $k \sim 0$, the dispersion relation is approximately linear close to the origin, and can fitted to extract the speed of sound. 
In the presence of disorder, rather than having a single linear branch the excitation spectrum instead takes the form of a continuum, however its lower boundary is still defined by a well-defined linear edge close to $k=0$. We find this edge by locating the lowest frequency at which the QSF reaches some fraction $\varepsilon=0.95$ of its maximum at each momentum, and we perform a linear fit of the points close to $k=0$ in order to obtain the sound velocity in the disordered superfluid.

In the gapped Mott insulator, there is no phonon branch in the dispersion relation, and the QSF carries no significant weight close to $\omega=0$. It therefore cannot be fitted with a linear slope in the low-$k$ regime. The speed of sound from this measure can be taken to be zero.
In the superfluid phase, the excitation spectrum displays a well-defined linear branch close to $k=0$, from which the speed of sound is extracted.
In the Bose glass phase, the system is gapless but insulating. Crucially, this means that there is no well-defined speed of sound, as phonons cannot propagate freely through the entire lattice. Rather than a linear slope close to $k=0$, the QSF instead exhibits a featureless continuum centered around $\omega=0$. In practice, we then set $V_{\rm{s}}=0$.

\begin{center}
\begin{figure}[h!]
\includegraphics[width=\linewidth]{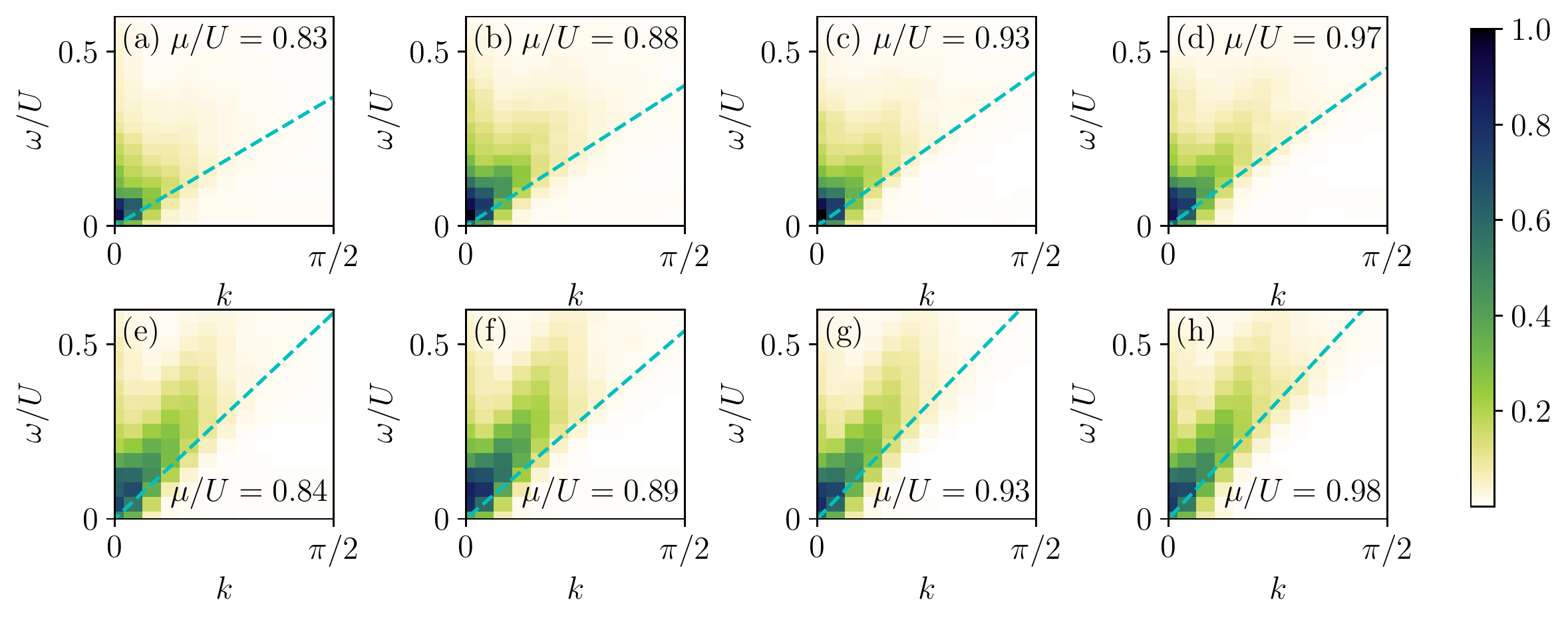}
\caption{QSF of $g_1(x,t)$, averaged over $N_{\rm{s}}=25$ disorder realizations, for four values of $\mu/U$ at two interaction strengths: (a-d) $J_{\rm f}/U=0.09$, and (e-h) $J_{\rm f}/U=0.12$. Cyan dashed lines indicate linear fits used to extract the speed of sound in the SF phase, the result of which is shown in Fig.~\ref{fig.speed_of_sound} of the main text using a threshold $\varepsilon=0.95$.
}
\label{fig.vfits}
\end{figure}
\end{center}

In Fig.~\ref{fig.vfits}, we show several examples of the fits used to obtain the results shown in Fig.~\ref{fig.speed_of_sound}. Note that many standard file readers apply automatic antialiasing to the image, resulting in blurring of the features. It is recommended to open this file in an alternate pdf viewer if this problem is encountered.

\clearpage
\subsection{Speed of sound in the homogeneous system}

\begin{center}
\begin{figure}[h!]
\includegraphics[width=0.65\linewidth]{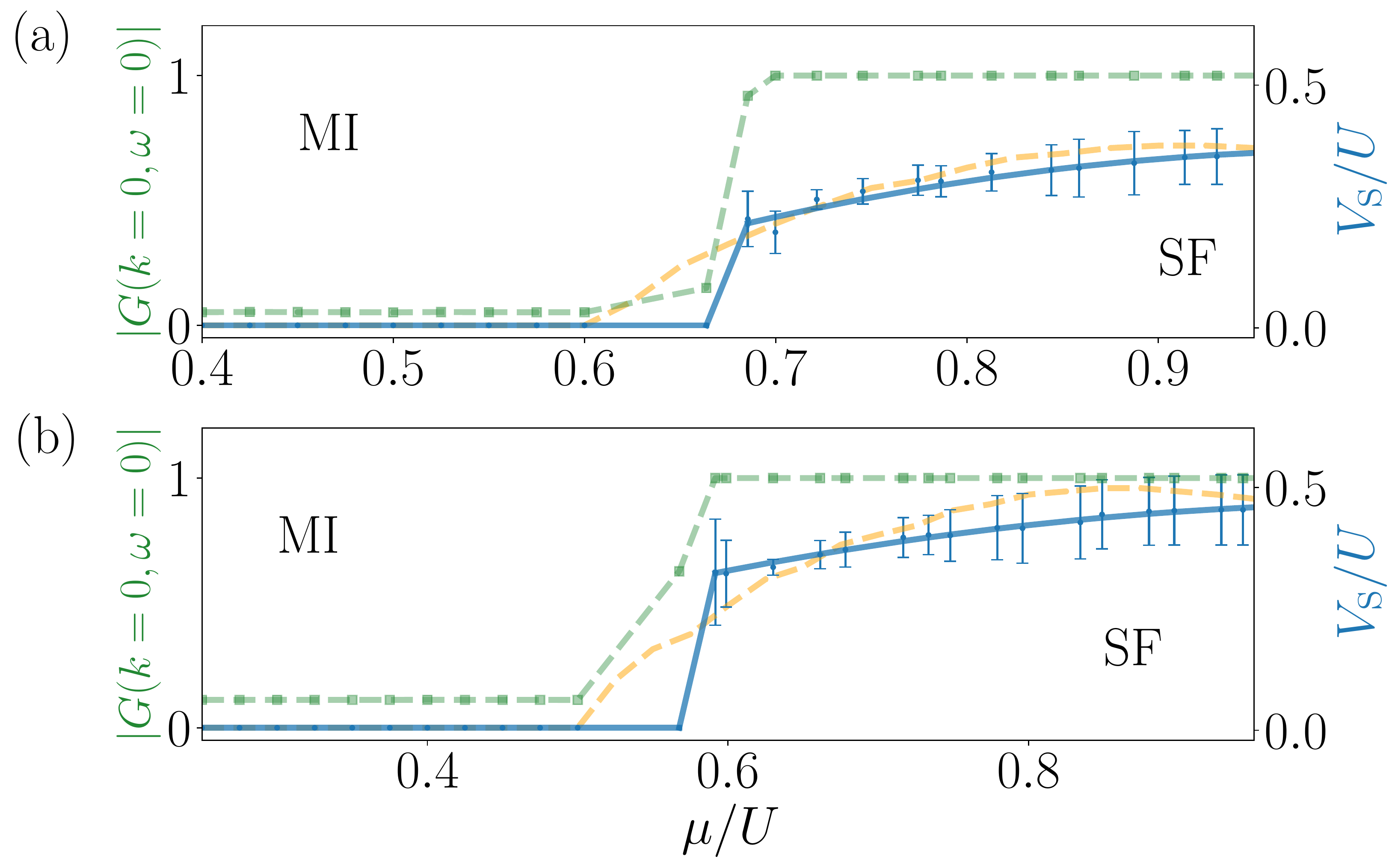}
\caption{\label{fig:fig_vs_clean}
Speed of sound resulting from fitting the QSF of $g_1(x,t)$ in the homogeneous system (no disorder) for (a)~$J_{\rm f}/U=0.09$ and (b) $J_{\rm f}/U=0.12$. The orange line is the result from the fermionic tight-binding model used in the main text for the disordered system. Note that very close to the MI-SF transition, it is difficult to accurately perform a linear fit in order to extract the speed of sound and so it is not possible to locate the transition point precisely. Away from the transition in the SF phase, however, the fermionized analytical prediction matches well with the numerical results.}
\end{figure}
\end{center}

For comparison with Fig.~\ref{fig.speed_of_sound} of the main text, we show in Fig.~\ref{fig:fig_vs_clean} the same quantities computed in the homogeneous (non-disordered) system, namely the speed of sound $V_{\rm{s}}$ and the amplitude of the gapless mode $|G(k=0,\omega=0)|$. Both are computed from the QSF of $g_1(x,t)$. We find that the sound velocity obtained from the homogeneous system in this regime is larger than the velocity reported in Fig.~\ref{fig.speed_of_sound} of the main text for the disordered system, and agrees with the analytical expression obtained in the strongly-interacting regime (within the error bars). This indicates that the deviation from the analytical result seen in the disordered system is mainly due to disorder suppressing the sound velocity, rather than from corrections of order $\sim J^2/U$, which would be present even in the clean system.

\clearpage
\subsection{Extracting $\xi/L$}

The spatially-resolved local spectral function (LSF) is defined as:
\begin{equation}
G(x,\omega) = 2\pi \sum_{n,n'} \rho_{\rm{i}}^{n'n} \delta(E_{n'} - E_{n} - \omega) \braket{n | \hat{O}(x) | n'}
\label{eq.LSF}
\end{equation}
where $\hat{O}(x)$ is some local observable. We do not compute the space-time Fourier transform as in the case of the QSF, but only the time-frequency transform. This allows us to obtain a spatially resolved measure of the excitation structure, and extract the lengthscale $\xi/L$ which characterizes the size of the gapped regions.

\begin{center}
\begin{figure}[h!]
\includegraphics[width=0.8\linewidth]{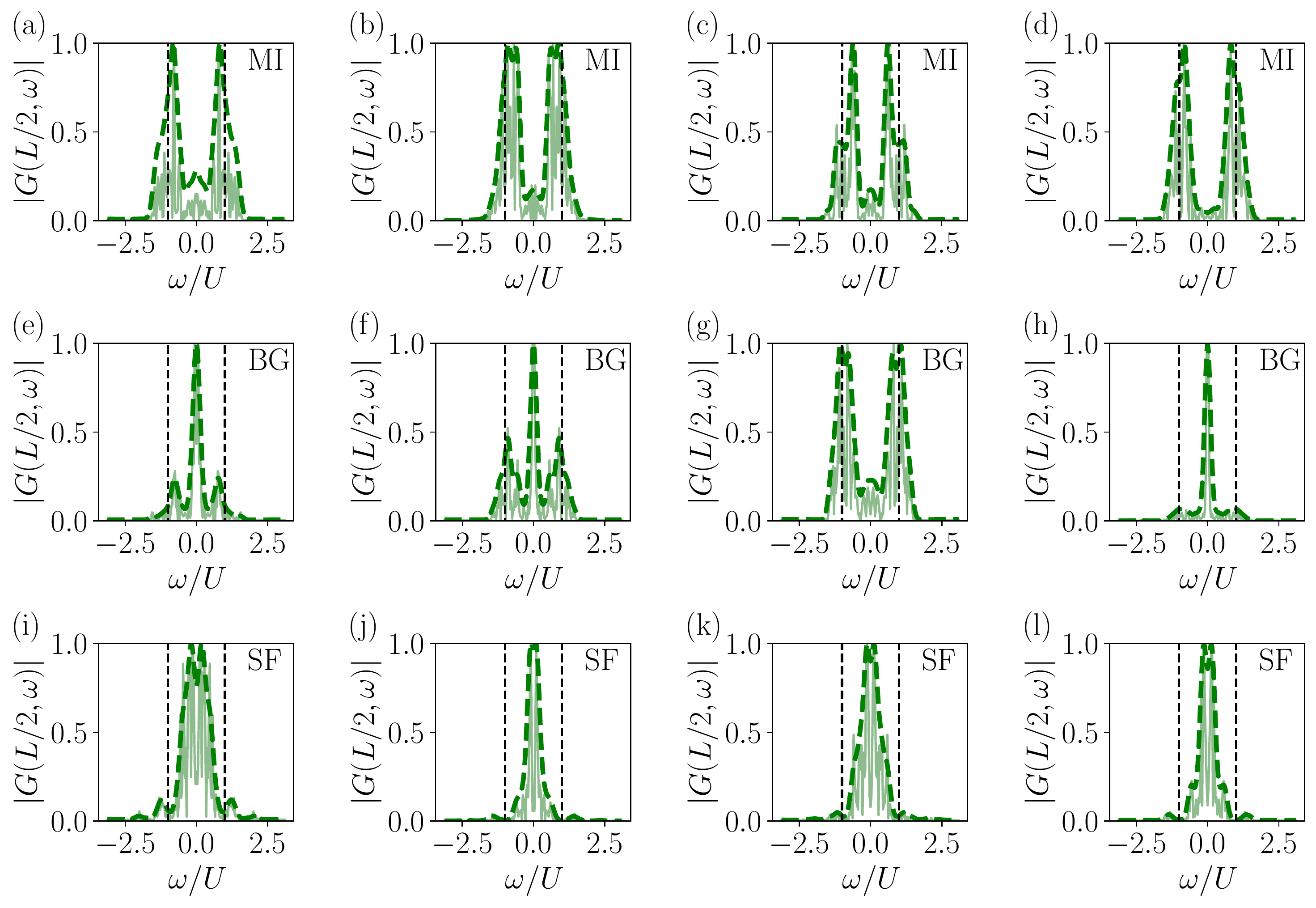}
\caption{A sample of normalized LSF for $J_{\rm{f}}/U=0.09$ and a range of different $\mu/U$ values, each taken on the central site of the chain and shown for a single disorder realization. The black dashed lines are guides to the eye located at $\omega/U=\pm1$. The chemical potentials used are
(a-d)~$\mu/U=0.5$ (MI phase),
(e-h)~$\mu/U=0.6$ (BG phase),
and (i-l)~$\mu/U=0.9$ (SF phase).
In the MI, the LSF is peaked at $\omega/U \approx \pm 1$, but the Gaussian convolution can cause a spurious large zero frequency signal, as seen in panel~(a). In the BG, the LSF for single realizations of the disordered potential can be peaked either at $\omega/U \approx 0$ [as in panels (e), (f), and (h)] or at $\omega/U \approx \pm 1$ [panel (g)], signifying the different forms of local order present in the BG phase in finite-size systems. In the SF, the LSF exhibits a strong peak at $\omega/U \approx 0$, consistent with the gapless excitations expected in this phase.}
\label{fig.threshold}
\end{figure}
\end{center}

As specified in the main text, we choose the observable to be the local density fluctuations, $\hat{O}(x)=\delta \hat{n}(x)$.
In Fig.~\ref{fig.threshold}, we show representative samples of the normalized LSF $G(x,\omega)$ taken on the central site of the chain $x=L/2$ in the Bose glass phase, for 10 different disorder realizations. We use a Gaussian convolution to smooth the LSF (dashed green lines) before extracting the weight of the $\omega=0$ contribution. Sites where $G(x,\omega=0) \geq \varepsilon$ (with $\varepsilon$ some threshold value) are defined as hosting gapless excitations, while sites with $G(x,\omega=0)<\varepsilon$ are identified as gapped. 

\vspace{-10mm}
In principle, one would expect a site to host gapless excitations for any $\varepsilon > 0$, however due to both the Gaussian convolution and the random disorder which modifies the density matrix coefficients in Eq.~(\ref{eq.LSF}) of the main text we must make a more conservative choice for the threshold $\varepsilon$. By convolving the signal with a broad Gaussian, we amplify the $\omega=0$ signal in the Mott insulator phase, and so a higher threshold of $\varepsilon > 0$ is required. In addition, there exist rare lattice sites in the MI phase where the peaks at $\omega/U = \pm 1$ are strongly suppressed due to the disorder, and are of the same order of magnitude as random noise in the signal. After normalizing the LSF, these random fluctuations can be amplified, and therefore it is again important to use a threshold $\varepsilon > 0$ that is larger than these random fluctuations. In practice, we find that varying the threshold in the range $0.25 < \varepsilon < 0.75$ does not strongly change the results except in the MI phase close to the tip of the Mott lobe (where the Mott gap closes exponentially). In this regime the gapped LSF after Gaussian convolution cannot be distinguished from true gapless excitations. In Fig.~\ref{fig.LSF} of the main text we choose $\varepsilon = 0.6$, based on systematic analysis of the maximum value of $G(R,\omega=0)$ for points known to be deep in the MI phase.

\end{document}